\documentclass[a4paper,11pt]{article}
\usepackage{pos}

\title{Neutrino Physics at Future Colliders}
\author*[a,b]{P. S. Bhupal Dev}

\affiliation[a]{Department of Physics and McDonnell Center for the Space Sciences, \\
Washington University, 
One Brookings Drive, St. Louis, MO 63130, USA}
\affiliation[b]{PRISMA$^+$ Cluster of Excellence \& Mainz Institute for Theoretical Physics, \\
Johannes Gutenberg-Universit\"{a}t Mainz, 55099 Mainz, Germany}

\emailAdd{bdev@wustl.edu}

\abstract{This is a brief review of the collider phenomenology of neutrino physics. Current and future colliders provide an ideal testing ground for (sub)TeV-scale neutrino
mass models, as they can directly probe the messenger particles, which could be either new fermions, scalars, or gauge bosons, associated with neutrino mass generation.
Moreover, the recent observation of TeV-scale neutrinos produced at the LHC offers new ways to test the limits of the Standard Model and beyond. }

\FullConference{Proceedings of the Corfu Summer Institute 2024 "School and Workshops on Elementary Particle Physics and Gravity" (CORFU2024)\
12 - 26 May, and 25 August - 27 September, 2024\\
Corfu, Greece\\}

\begin{document}
\maketitle

\section{Introduction}
The Standard Model (SM), a cornerstone of particle physics, is considered the most empirically successful theory in the history of science. It has made remarkably accurate predictions about fundamental particle properties and interactions, which have been experimentally confirmed with high precision. Nonetheless, there exist certain phenomena which cannot be explained within the framework of SM, and therefore, necessitate the existence of beyond-the-SM (BSM) physics. The neutrino sector provides a prime example. On one hand, neutrinos have played a key role in establishing two of the most important features of the SM, namely, (i) maximal parity violation in weak interactions, and (ii) 3-fold repetition of the fermion family structure. On the other hand, the observation of neutrino oscillations implies that the neutrino flavor eigenstates are a linear superposition of their mass eigenstates, and at least two of the mass eigenvalues must be nonzero in order to explain the observed solar and atmospheric mass-squared differences. This is a clear evidence of BSM physics, because neutrinos are predicted to be massless in the SM. Therefore, a better understanding of the neutrino sector will shed some light on the associated BSM physics. In this proceeding\footnote{See \url{https://indico.cern.ch/event/1349196/contributions/5833331/} for the talk slides.}, we review how current and future colliders at the energy frontier can play a crucial role, complementary to dedicated neutrino experiments at the intensity frontier, in advancing our knowledge of the neutrino sector. 

\section{`Seeing' Neutrinos at Colliders}
Under collider environments, neutrinos simply behave as missing energy, and therefore, are generally undetectable. But it was noticed long ago that the LHC produces a large number of neutrinos~\cite{DeRujula:1984ns}, strongly collimated along the beam collision axis. As the LHC tunnel eventually curves away, these neutrinos can be accessed by placing a detector further downstream from the interaction point. This is the idea behind dedicated collider neutrino experiments like FASER and SND@LHC, both of which recently reported the first observation of collider neutrinos~\cite{FASER:2023zcr, SNDLHC:2023pun}. A continuation and expansion of this promising collider neutrino experimental program during the HL-LHC era is anticipated at the multi-purpose Forward Physics Facility (FPF)~\cite{Feng:2022inv}. 

Since collider neutrinos mainly come from the decay of charged mesons in the forward region, neutrino flux measurements at FPF provide a novel way to probe forward hadron production, offering unique insights into the strong interaction dynamics in previously inaccessible kinematic regions and validating hadron interaction models~\cite{Fieg:2023kld}. The collider neutrino experiments also enable the measurement of neutrino interaction cross sections at previously inaccessible TeV energies~\cite{FASER:2024hoe}. Future measurements of all three neutrino flavors offer a unique opportunity to test lepton universality in neutrino scattering. High-statistics neutrino scattering data from forward neutrino experiments can also significantly reduce the proton and nuclear parton distribution function (PDF) uncertainties~\cite{Cruz-Martinez:2023sdv}, which will benefit key measurements at the LHC, for both SM and BSM physics. Another important aspect of collider neutrino experiments like FASER$\nu$ is that the emulsion detectors have excellent spatial resolution, at the $\mu$m level or below, allowing for precision studies of tau neutrino properties and the first separate observation of tau neutrinos and tau antineutrinos. This is especially important for rare SM events like tau tridents~\cite{Altmannshofer:2024hqd, Bigaran:2024zxk}. Although we expect only a handful of tau tridents even at FASER$\nu$2, a future neutrino factory at the proposed muon collider will produce thousands of tau tridents~\cite{Bojorquez-Lopez:2024bsr} which can be used to do precision SM physics, as well as BSM physics searches involving the least-studied tau sector.

\section{Nature of Neutrino Mass}
The nature of neutrino mass remains a fundamental open question. In the SM, neutrinos are purely left-handed ($\nu_L$). If we add the right-handed partners ($N$), they can acquire a Dirac mass term via the Higgs mechanism, just like the other SM fermions do. However, as neutrinos are the only neutral fermions in the SM, there is another  possibility for them to acquire a Majorana mass, if the lepton number symmetry of the SM is somehow broken. The key question is how to distinguish a Dirac neutrino from a Majorana  one  experimentally.  Any such distinction is expected to be suppressed by some power of $m_\nu/E_\nu$ (where $m_\nu$ and $E_\nu$ are the neutrino mass and energy), which makes it unobservable in most experimental settings with relativistic neutrinos. A direct detection of the (non-relativistic) cosmic relic neutrino background can in principle address this issue~\cite{Long:2014zva, PTOLEMY:2019hkd}; however, the feasibility of such experiments remains questionable~\cite{PTOLEMY:2022ldz}.

A `smoking gun' signal of Majorana neutrinos would be the observation of a process with lepton number violation (LNV), such as the 
neutrinoless double beta decay ($0\nu\beta\beta$)~\cite{Agostini:2022zub}. Currently, there exists only a lower limit on the $0\nu\beta\beta$ lifetime: $T_{1/2}^{0\nu}\gtrsim 10^{26}$~yr, corresponding to an upper limit on the effective Majorana mass $m_{\beta\beta}\lesssim 0.1$ eV~\cite{GERDA:2020xhi, KamLAND-Zen:2022tow}. 
The next-generation ton-scale experiments like LEGEND-1000~\cite{LEGEND:2021bnm} and nEXO~\cite{nEXO:2021ujk} are expected to reach a half-life sensitivity of ${\cal O}(10^{28}~\textrm{yr})$, corresponding to $m_{\beta\beta}\lesssim 0.02$ eV. However, there is no guarantee that a positive $0\nu\beta\beta$ signal will be ever observed, especially if neutrino masses turn out to follow normal ordering, which is, in fact, preferred over the inverted ordering in current global oscillation fits~\cite{Esteban:2024eli}, as well as from stringent cosmological constraints on the sum of neutrino masses~\cite{DESI:2024mwx}. 

High-energy colliders provide an alternative probe of the Majorana nature of neutrinos~\cite{Deppisch:2015qwa, Cai:2017mow}, complementary to the low-energy probes like $0\nu\beta\beta$. The basic idea is to search for the signatures of the messengers of neutrino mass which might encode the information on LNV in the underlying theory. In the SM, neutrinos are massless because of the accidental global $B-L$ symmetry. Therefore, the simplest way to generate neutrino masses is by breaking the $B-L$, usually parameterized by the dimension-5 Weinberg operator $LLHH/\Lambda$~\cite{Weinberg:1979sa}, where $L$ and $H$ are the $SU(2)_L$ lepton and Higgs doublets respectively, and $\Lambda$ is the $(B-L)$-breaking scale. There are only three tree-level realizations of this operator, commonly known as the type-I~\cite{Minkowski:1977sc, Mohapatra:1979ia, Yanagida:1979as, Gell-Mann:1979vob}, II~\cite{Schechter:1980gr, Magg:1980ut, Cheng:1980qt,Lazarides:1980nt, Mohapatra:1980yp}, and III~\cite{Foot:1988aq} seesaw mechanisms. Searching for the corresponding messenger particles, namely, $SU(2)_L$-singlet fermions, $SU(2)_L$-triplet scalars and fermions respectively,  at the high-energy colliders could provide a powerful test of LNV, and hence, the Majorana nature of neutrino masses~\cite{Keung:1983uu, Deppisch:2015qwa, Cai:2017mow}.        

\section{Sterile Neutrinos}
Perhaps the simplest possibility to generate neutrino mass is by adding SM-singlet, right-handed neutrinos $N$ (also known as sterile neutrinos or heavy neutral leptons), which enables us to write not only the Dirac mass term $m_D\bar{\nu}_LN$+h.c., but also a Majorana mass term $m_N\bar{N}^cN$, which leads to the neutrino mass matrix
\begin{align}
  M_\nu = \begin{pmatrix}
        0 & m_D \\ m_D^T & m_N
    \end{pmatrix}.
\end{align}
In a bottom-up phenomenological approach, the scale $m_N$ is unknown. If $||m_N||\gg ||m_D||$, the light neutrino mass matrix becomes $m_\nu\simeq -m_D m_N^{-1}m_D^T$, as in the original type-I seesaw formulation motivated by $SO(10)$ GUT~\cite{Mohapatra:1979ia}, and neutrinos are purely Majorana particles. However, the observability of LNV very much depends on the sterile neutrino properties. 
Experimentally, the most relevant quantity for sterile neutrino searches is the active-sterile mixing $V_{\ell N}$, along with the scale of $m_N$. In the naive seesaw limit, they are related: $V_{\ell N}\simeq m_D m_N^{-1}\simeq (m_\nu m_N^{-1})^{1/2}$. However, this relation is strictly valid only in the single-generation case. With more than one generation,\footnote{
We need at least two sterile species to explain the solar and atmospheric
mass-squared differences by tree-level type-I seesaw mechanism, and often three to cancel anomalies in UV-complete models.} `large' mixing is allowed even for low-scale $m_N$, while being consistent with the seesaw formula, by choosing special textures of $m_D$ and $m_N$~\cite{Kersten:2007vk}. Variations of the type-I seesaw, such as inverse~\cite{Mohapatra:1986bd}, linear~\cite{Malinsky:2005bi}, or radiative~\cite{Dev:2012sg} seesaw with additional SM-singlet particles, can also realize such low-scale seesaw with `large' mixing in a technically natural way. 

\begin{figure}[t!]
  %  \vspace{-0.5cm} 
  \includegraphics[width=1.0\linewidth]{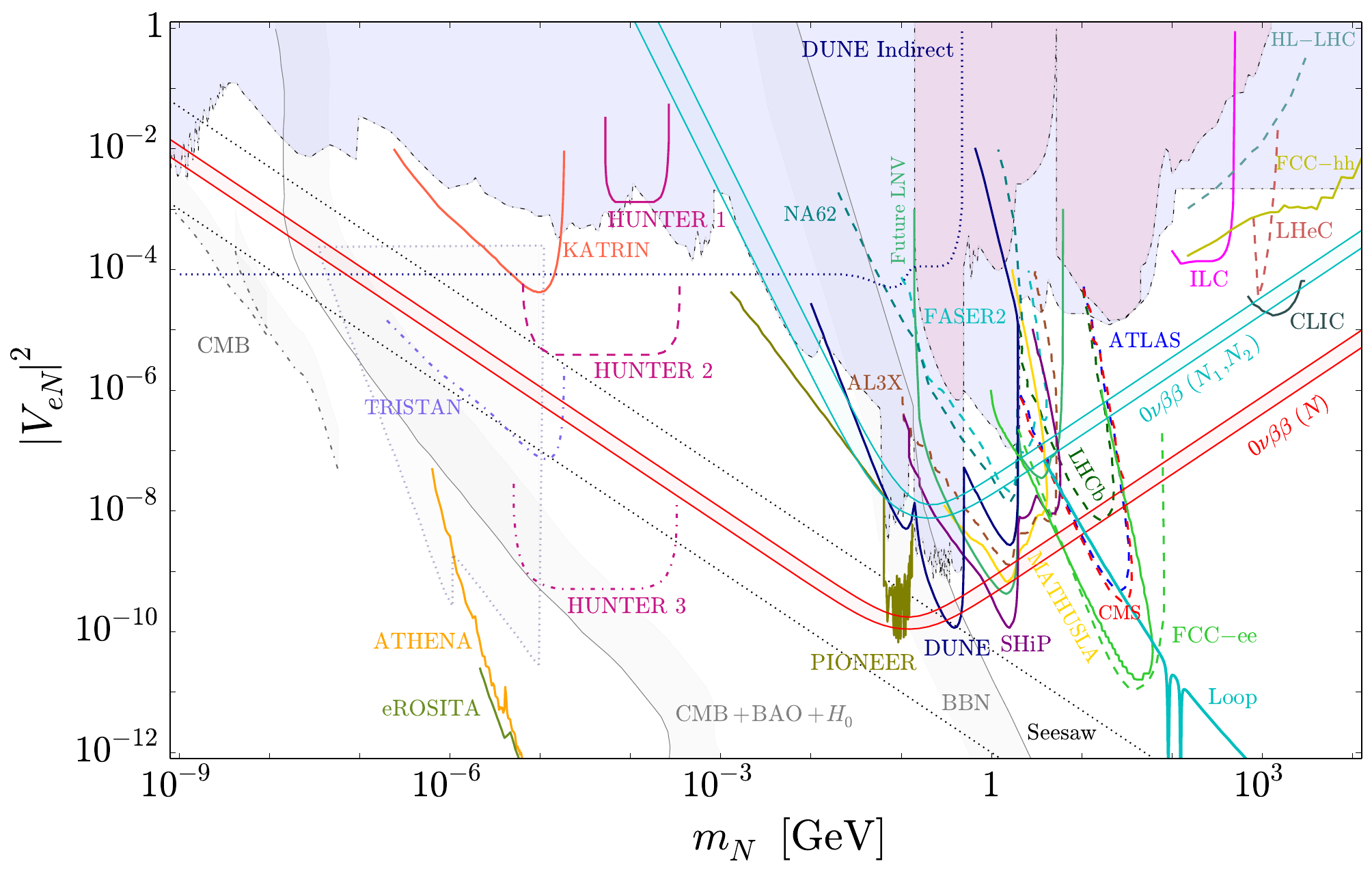}
   % \vspace{-0.8cm} 
    \caption{Current constraints (shaded regions) and future projections (open curves) on the active-sterile mixing strength in the electron flavor as a function of the sterile neutrino mass. The region in light blue is disallowed for both Dirac and
Majorana sterile neutrinos whereas the region in light red applies only for Majorana neutrinos. The $0\nu\beta\beta$ 
sensitivities are for a heavy sterile Majorana (red) and quasi-Dirac neutrino (teal) for the half-life $T_{1/2}^{0\nu}=10^{28}$ yr in $^{76}$Ge. Adapted from Refs.~\cite{Bolton:2019pcu, Bolton:2022pyf}.}
\label{fig:sterile}
%\vspace{-0.8 cm} 
\end{figure}

Without specifying any explicit model constructions, we can treat the sterile neutrino mass and mixing as free parameters in a bottom-up phenomenological approach. This is the approach usually taken by the experimental searches of sterile neutrinos and exclusion limits are derived in the sterile neutrino mass-mixing plane assuming a single flavor dominance.\footnote{Reinterpreting these limits in realistic models to reproduce the neutrino oscillation data often result in weaker constraints, depending on the model details~\cite{Tastet:2021vwp}.}
A summary of the existing constraints and future prospects in the sterile neutrino mass and mixing plane is shown in Fig.~\ref{fig:sterile}. Here we only show the mixing with electron flavor; similar plots for the muon and tau flavors, as well as future experimental prospects, can be found,~e.g.~in Refs.~\cite{Bolton:2019pcu, Antusch:2016ejd, Abdullahi:2022jlv}. The entire mass range shown here can be divided into 4 sectors, depending on which type of constraint is dominant: (i) astrophysical and cosmological constraints in the sub-MeV range, (ii) beam dump and meson decay constraints in the MeV--GeV range; (iii) collider constraints in the GeV--TeV range; and (iv) electroweak precision  constraints beyond TeV. Note that in the electron mixing case, although the $0\nu\beta\beta$ constraint is usually the most stringent, it can be significantly weakened depending on the sterile mass hierarchy and the relative phases between them~\cite{Bolton:2019pcu}, without affecting most of the other laboratory constraints. Similarly, the cosmological constraints, especially from BBN and CMB, could in principle be weakened in presence of additional (dark) interactions of the sterile neutrino~\cite{Dasgupta:2021ies}.  Therefore, it is important to search for the sterile neutrinos in as many independent ways as possible, even if the projected sensitivities in some cases may not be as competitive as the existing constraints.  

At the high-energy colliders, the heavy sterile neutrinos are produced in charged-current and neutral-current processes through their admixture with the active states: $pp\to W^+\to \ell^+N$ and $pp/e^+e^-\to Z\to \nu N$. If kinematically allowed, $pp\to h\to \nu (N) N$ and $e^+e^-\to Zh\to Z\nu (N) N$ also provide another promising probe of sterile neutrinos using precision Higgs data~\cite{Dev:2012zg, Das:2017zjc, Gao:2021one}. The collider constraints in the GeV-TeV range are of two types, depending on whether the sterile neutrino decays promptly or is long-lived. The former is the case for heavier sterile neutrinos with relatively larger mixing, while the latter is applicable for lighter ones with smaller mixing. At the LHC, both prompt~\cite{ATLAS:2019kpx, CMS:2024xdq} and displaced~\cite{CMS:2024hik,ATLAS:2025uah} vertex searches for sterile neutrinos have been performed and bounds have been derived as shown in Fig.~\ref{fig:sterile}. The future colliders are expected to extend the sensitivity reach significantly. In particular, future lepton colliders like FCC-ee or CEPC running at the $Z$-pole are especially sensitive to the long-lived sterile neutrinos~\cite{Blondel:2014bra}, and are capable of reaching all the way to the theoretical seesaw limit. For a detailed discussion of the future collider sensitivities to the sterile neutrino parameter space, see e.g.~Refs.~\cite{Li:2023tbx,Bellagamba:2025xpd} and references therein.   

As for the LNV signals originating from sterile neutrinos at colliders, the `smoking gun' signal is the so-called Keung-Senjanovi\'{c} process: $pp\to N\ell^\pm\to \ell^\pm \ell^\pm jj$ with no missing transverse energy~\cite{Keung:1983uu}. However, in the minimal scenario with no additional interactions beyond the active-sterile neutrino mixing, the LNV signal is expected to be suppressed by the light neutrino mass~\cite{Fernandez-Martinez:2015hxa}, since in the large mixing limit required for a sizable production cross-section, the sterile neutrinos are pseudo-Dirac~\cite{Kersten:2007vk}. There are two notable exceptions to this common lore, even in the minimal scenario: (i) When the sterile mass splitting $\Delta m_N$ is of the same order as its decay width $\Gamma_N$, when the LNV amplitude is resonantly enhanced~\cite{Bray:2007ru, Anamiati:2016uxp}; and (ii) Sterile neutrino-antineutrino oscillations, which can be resolved in long-lived particle searches at future colliders~\cite{Antusch:2017ebe,Antusch:2024otj}.     

\section{Going Beyond the Minimal Scenario}
In the presence of additional gauge interactions, the collider phenomenology of sterile neutrinos gets enriched. This is because the sterile neutrinos can now be produced directly via the gauge interactions without the active-sterile mixing suppression, but instead only suppressed by the heavy gauge boson mass. For example, in the $U(1)'$ extension with a $Z'$ boson coupling to sterile neutrinos, there is a new production channel: $pp/e^+e^-\to Z'\to NN$~\cite{Deppisch:2013cya,Das:2019fee, Das:2021esm}, and in fact, the $Z'\to NN$ branching ratio can be enhanced with respect to $Z'\to$SM for certain $U(1)'$ charge assignments~\cite{Das:2017flq}. Moreover, future lepton colliders offer a great opportunity to probe the leptophilic $U(1)_{L_\alpha-L_\beta}$ models~\cite{Dasgupta:2023zrh}, which are not so well-constrained by the LHC. Similarly, in the $SU(2)_L\times SU(2)_R\times U(1)_{B-L}$ extension, the sterile neutrinos can be produced via the right-handed gauge current: $pp\to W_R\to N\ell_R^\pm$~\cite{Keung:1983uu, Chen:2013foz}, thus enhancing the collider sensitivity to LNV signals~\cite{Nemevsek:2018bbt}. Moreover, the longitudinal beam polarization option at a future $e^+e^-$ collider could provide an unambiguous distinction between the left- and right-handed contributions to the sterile neutrino production~\cite{Biswal:2017nfl}.  Note that in both $U(1)_{B-L}$ and left-right gauge extensions, the addition of three sterile neutrinos is essential to cancel the anomalies, thus providing a strong theoretical motivation for their existence and for the UV-completion of the seesaw mechanism.  

Apart from the type-I seesaw and its variants, there are other ways to realize the dimension-5 Weinberg operator for LNV. For instance, in the type-II~\cite{Schechter:1980gr, Magg:1980ut, Cheng:1980qt,Lazarides:1980nt, Mohapatra:1980yp}  seesaw, one introduces $SU(2)_L$-triplet scalars $\Delta_L=(\Delta^{++},\Delta^+,\Delta^0)_L$, which can be directly produced at colliders via their $SU(2)_L$ gauge interactions. In this case, the `smoking gun' signal would be the detection of the doubly-charged scalar with LNV interactions: $pp/e^+e^-\to \gamma/Z\to \Delta^{++}\Delta^{--}$, followed by the decays $\Delta^{\pm\pm}\to \ell^\pm \ell^\pm$ or $W^\pm W^\pm$ depending on the value of the triplet vacuum expectation value~\cite{FileviezPerez:2008jbu}. Both dilepton and $WW$ decay channels of the doubly-charged scalar have been searched for at the LHC and lower limits have been set on the doubly-charged scalar mass~\cite{CMS:2017pet, ATLAS:2021jol, ATLAS:2022pbd}. In the minimal left-right model, there also exists an $SU(2)_R$-triplet scalar $\Delta_R=(\Delta^{++},\Delta^+,\Delta^0)_R$, which leads to additional collider phenomenology~\cite{Gunion:1989in, Dev:2016dja}. In particular, the neutral component $\Delta_R^0$ is hadrophobic and is allowed to be very light, thus leading to novel displaced vertex signals at future colliders~\cite{Dev:2016nfr, Dev:2017dui}. Similarly, the single and pair production of $\Delta_R$ via its mixing with the SM Higgs boson leads to novel LNV signals at colliders~\cite{Maiezza:2015lza, Nemevsek:2016enw}.

\section{Radiative Neutrino Mass}
An alternative to the tree-level seesaw for small neutrino masses is that they arise only as quantum corrections. In these radiative neutrino mass models~\cite{Cai:2017jrq},
the tree-level Lagrangian does not generate the dimension-5 LNV operator, owing to the particle content or
symmetries present in the model, but small
Majorana neutrino masses are induced only at the loop level. Some well-known examples are the Zee model~\cite{Zee:1980ai} at one-loop, Zee-Babu model~\cite{Zee:1985id,Babu:1988ki} at two-loop, and KNT model~\cite{Krauss:2002px} at three-loop order. All these models involve heavy particles coupling to the SM going in the loops, which lead to potentially observable collider signals. 

\begin{figure}[!t]
\includegraphics[width=0.99\textwidth]{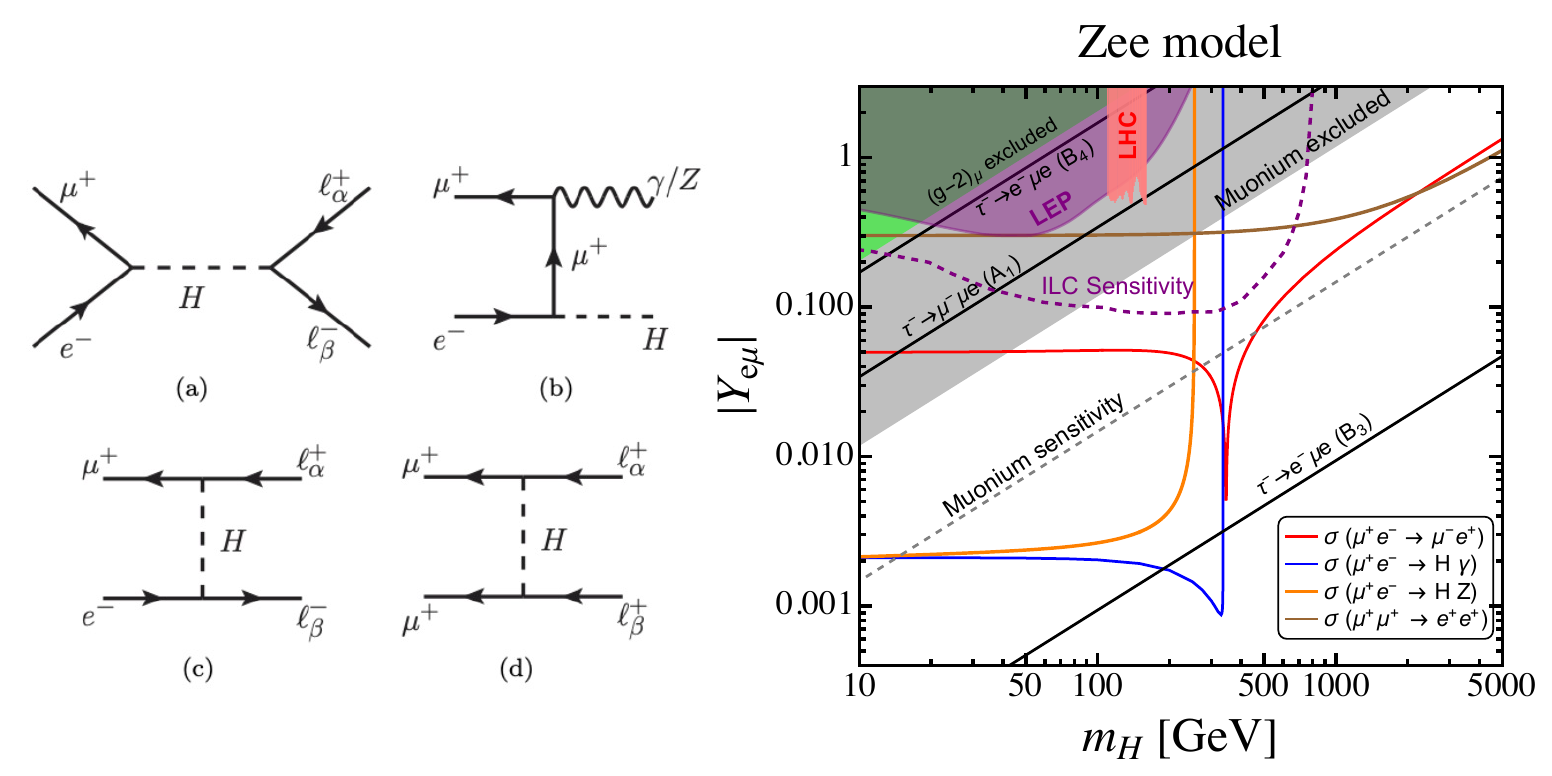}
    \caption{{\it Left:} Relevant Feynman diagrams for the processes involving the neutral scalar $H$ in the Zee model  at $\mu$TRISTAN. {\it Right:} $\mu$TRISTAN sensitivity to the Zee model parameter space for various channels as shown in the left panel.  The shaded regions are excluded: Purple (pink) shaded from LEP (LHC) dilepton data, green shaded from $(g-2)_\mu$, and gray shaded from muonium oscillation. The future muonium and ILC sensitivities are shown by the black and purple dashed lines, respectively. The solid black lines show the $\tau$ LFV constraints for different Yukawa textures. Adapted from Ref.~\cite{Dev:2023nha}. 
   } 
    \label{fig:Zeeplot}
\end{figure}

Take the Zee model as a prototypical example. It features an additional Higgs doublet and a charged scalar singlet, both assumed to be leptophilic in nature. The physical mass eigenstates for the charged scalars are constrained to be roughly above 100 GeV from LEP searches~\cite{ALEPH:2013htx, Babu:2019mfe}. As for the extra neutral scalars, they must have lepton flavor violating (LFV) couplings to fit neutrino oscillation data. Thus, their Yukawa couplings can have stringent charged LFV constraints, depending on the Yukawa texture. Nevertheless, future lepton colliders like $\mu$TRISTAN could still probe a huge swath of the allowed parameter space~\cite{Dev:2023nha}, as shown in Fig.~\ref{fig:Zeeplot}. 

In addition to having the collider tests, the radiative neutrino mass models with at least one SM particle in the loop generically lead to observable neutrino non-standard interactions~\cite{Babu:2019mfe}, thus making them testable in the future neutrino oscillation experiments as well.   

\section{LHC as a Lepton Collider}
Due to quantum fluctuations, protons also contain charged leptons, making it possible to study lepton-induced processes at the LHC as well.
The simplest process of this kind consists of the collision between a lepton from one proton and a quark  from the
other proton, giving rise to the resonant production of a single leptoquark (LQ) state~\cite{Buonocore:2020erb}. This is complementary to quark- or
gluon-induced LQ production channels. The LQs could play an important role in neutrino mass generation via radiative mechanism~\cite{Babu:2019mfe}, and therefore, directly probing the LQ-lepton-quark coupling using future colliders is a promising way to test these radiative models. 

The lepton PDFs of the proton also allows us to use the LHC effectively as a lepton collider (sans the cleaner environment of actual lepton colliders), in the sense that even pure leptophilic particles can be resonantly produced at the LHC. One such example is the resonant production of leptophilic neutral Higgs bosons in the Zee model at the LHC~\cite{Afik:2023vyl}.

\section{Connection to Leptogenesis and Dark Matter}
Neutrinos can also provide the missing link to other BSM phenomena. For example, the same sterile neutrinos responsible for neutrino mass could also explain the observed baryon asymmetry of the Universe via the mechanism of leptogenesis~\cite{Fukugita:1986hr, Bodeker:2020ghk}. Although the vanilla leptogenesis is a high-scale mechanism inaccessible to colliders, there exist low-scale variants, such as   the resonant (or freeze-out) leptogenesis~\cite{Pilaftsis:2003gt} and ARS (or freeze-in) leptogenesis~\cite{Akhmedov:1998qx} that can be directly tested at colliders. In fact, it was recently shown that there is a smooth transition from the freeze-in regime to the freeze-out regime~\cite{Klaric:2020phc}. It was also shown~\cite{Drewes:2021nqr} that the range of mixings for which leptogenesis is feasible in the seesaw model with three sterile neutrinos is considerably larger
than in the minimal model with only two heavy neutrinos and extends all the way up to the current
experimental bounds. For such large mixing angles the HL-LHC could potentially observe a number
of events that is large enough to compare different decay channels, which would be a first step towards testing the connection between the origin of matter and neutrino masses. 

Similarly, the lightest sterile neutrino, if cosmologically stable and in the keV mass regime, can be a good (warm) DM candidate~\cite{Drewes:2016upu}. This is in contrast with the active neutrinos in the SM which cannot explain the observed DM relic density, because they decouple as hot relics. The minimal version of the model where sterile neutrino DM production happens solely from active-sterile oscillations -- the so-called Dodelson-Widrow mechanism~\cite{Dodelson:1993je}, is now ruled out by $X$-ray line searches. However, extensions of the model featuring additional neutrino (self) interactions either in the active or in the sterile sector can still have a viable sterile neutrino production mechanism that is consistent with the current bounds~\cite{DeGouvea:2019wpf, Astros:2023xhe}.   

Even in the minimal left-right symmetric extension, keV-scale lightest right-handed neutrino remains as a good warm DM candidate~\cite{Nemevsek:2012cd}. In addition, the neutral component of the $SU(2)_R$-triplet scalar field can be an alternative decaying DM candidate~\cite{Dev:2025fcv}. 

The best signature of these keV-scale DM candidates is a nearly monochromatic line feature in $X$-ray and gamma-ray observations~\cite{Abazajian:2017tcc}. The future colliders can complement the astrophysical signatures with the standard monophoton or monojet signals, which are equally applicable to light (keV-scale) DM.

\section{Conclusion}
Neutrinos have given us the first (and so far only) laboratory evidence of BSM physics. Even in the precision era of neutrino oscillation physics, there are many open questions in neutrino theory. To address them, it is necessary to cast a broad net covering multiple experimental frontiers and exploring potential connections to other BSM puzzles, 
including baryon asymmetry and dark matter. From directly `seeing' neutrinos to probing the neutrino mass mechanism, high-energy future colliders provide an  unprecedented opportunity to study  neutrinos -- the least known sector of the SM.

\section*{Acknowledgments}
The work of B.D. was partly supported by the U.S. Department of Energy under grant No. DE-SC0017987, and by a Humboldt Fellowship from the Alexander von Humboldt Foundation. 
\bibliographystyle{utphys}
\bibliography{ref_dev.bib}

\providecommand{\href}[2]{#2}\begingroup\raggedright\begin{thebibliography}{100}

\bibitem{DeRujula:1984ns}
A.~De~Rujula, \href{http://dx.doi.org/https://inis.iaea.org/records/w2mm7-mt076}{``{Neutrino Physics at Future Colliders},''}
\newblock 1984.
\newblock Proceedings of the 6th General Conference of the EPS ``Trends in Physics'', Vol. 1, 236-245.

\bibitem{FASER:2023zcr}
{\bfseries FASER} Collaboration, H.~Abreu {\em et~al.}, ``{First Direct Observation of Collider Neutrinos with FASER at the LHC},'' \href{http://dx.doi.org/10.1103/PhysRevLett.131.031801}{{\em Phys. Rev. Lett.} {\bfseries 131} no.~3, (2023) 031801}, \href{http://arxiv.org/abs/2303.14185}{{\ttfamily arXiv:2303.14185 [hep-ex]}}.

\bibitem{SNDLHC:2023pun}
{\bfseries SND@LHC} Collaboration, R.~Albanese {\em et~al.}, ``{Observation of Collider Muon Neutrinos with the SND@LHC Experiment},'' \href{http://dx.doi.org/10.1103/PhysRevLett.131.031802}{{\em Phys. Rev. Lett.} {\bfseries 131} no.~3, (2023) 031802}, \href{http://arxiv.org/abs/2305.09383}{{\ttfamily arXiv:2305.09383 [hep-ex]}}.

\bibitem{Feng:2022inv}
J.~L. Feng {\em et~al.}, ``{The Forward Physics Facility at the High-Luminosity LHC},'' \href{http://dx.doi.org/10.1088/1361-6471/ac865e}{{\em J. Phys. G} {\bfseries 50} no.~3, (2023) 030501}, \href{http://arxiv.org/abs/2203.05090}{{\ttfamily arXiv:2203.05090 [hep-ex]}}.

\bibitem{Fieg:2023kld}
M.~Fieg, F.~Kling, H.~Schulz, and T.~Sj\"ostrand, ``{Tuning pythia for forward physics experiments},'' \href{http://dx.doi.org/10.1103/PhysRevD.109.016010}{{\em Phys. Rev. D} {\bfseries 109} no.~1, (2024) 016010}, \href{http://arxiv.org/abs/2309.08604}{{\ttfamily arXiv:2309.08604 [hep-ph]}}.

\bibitem{FASER:2024hoe}
{\bfseries FASER} Collaboration, R.~Mammen~Abraham {\em et~al.}, ``{First Measurement of \ensuremath{\nu}e and \ensuremath{\nu}\ensuremath{\mu} Interaction Cross Sections at the LHC with FASER\textquoteright{}s Emulsion Detector},'' \href{http://dx.doi.org/10.1103/PhysRevLett.133.021802}{{\em Phys. Rev. Lett.} {\bfseries 133} no.~2, (2024) 021802}, \href{http://arxiv.org/abs/2403.12520}{{\ttfamily arXiv:2403.12520 [hep-ex]}}.

\bibitem{Cruz-Martinez:2023sdv}
J.~M. Cruz-Martinez, M.~Fieg, T.~Giani, P.~Krack, T.~M\"akel\"a, T.~R. Rabemananjara, and J.~Rojo, ``{The LHC as a Neutrino-Ion Collider},'' \href{http://dx.doi.org/10.1140/epjc/s10052-024-12665-1}{{\em Eur. Phys. J. C} {\bfseries 84} no.~4, (2024) 369}, \href{http://arxiv.org/abs/2309.09581}{{\ttfamily arXiv:2309.09581 [hep-ph]}}.

\bibitem{Altmannshofer:2024hqd}
W.~Altmannshofer, T.~M\"akel\"a, S.~Sarkar, S.~Trojanowski, K.~Xie, and B.~Zhou, ``{Discovering neutrino tridents at the Large Hadron Collider},'' \href{http://dx.doi.org/10.1103/PhysRevD.110.072018}{{\em Phys. Rev. D} {\bfseries 110} no.~7, (2024) 072018}, \href{http://arxiv.org/abs/2406.16803}{{\ttfamily arXiv:2406.16803 [hep-ph]}}.

\bibitem{Bigaran:2024zxk}
I.~Bigaran, P.~S.~B. Dev, D.~Lopez~Gutierrez, and P.~A.~N. Machado, ``{Tau Tridents at Accelerator Neutrino Facilities},'' \href{http://arxiv.org/abs/2406.20067}{{\ttfamily arXiv:2406.20067 [hep-ph]}}.

\bibitem{Bojorquez-Lopez:2024bsr}
L.~Bojorquez-Lopez, M.~Hostert, C.~A. Arg\"uelles, and Z.~Liu, ``{The Neutrino Slice at Muon Colliders},'' \href{http://arxiv.org/abs/2412.14115}{{\ttfamily arXiv:2412.14115 [hep-ph]}}.

\bibitem{Long:2014zva}
A.~J. Long, C.~Lunardini, and E.~Sabancilar, ``{Detecting non-relativistic cosmic neutrinos by capture on tritium: phenomenology and physics potential},'' \href{http://dx.doi.org/10.1088/1475-7516/2014/08/038}{{\em JCAP} {\bfseries 08} (2014) 038}, \href{http://arxiv.org/abs/1405.7654}{{\ttfamily arXiv:1405.7654 [hep-ph]}}.

\bibitem{PTOLEMY:2019hkd}
{\bfseries PTOLEMY} Collaboration, M.~G. Betti {\em et~al.}, ``{Neutrino physics with the PTOLEMY project: active neutrino properties and the light sterile case},'' \href{http://dx.doi.org/10.1088/1475-7516/2019/07/047}{{\em JCAP} {\bfseries 07} (2019) 047}, \href{http://arxiv.org/abs/1902.05508}{{\ttfamily arXiv:1902.05508 [astro-ph.CO]}}.

\bibitem{PTOLEMY:2022ldz}
{\bfseries PTOLEMY} Collaboration, A.~Apponi {\em et~al.}, ``{Heisenberg\textquoteright{}s uncertainty principle in the PTOLEMY project: A theory update},'' \href{http://dx.doi.org/10.1103/PhysRevD.106.053002}{{\em Phys. Rev. D} {\bfseries 106} no.~5, (2022) 053002}, \href{http://arxiv.org/abs/2203.11228}{{\ttfamily arXiv:2203.11228 [hep-ph]}}.

\bibitem{Agostini:2022zub}
M.~Agostini, G.~Benato, J.~A. Detwiler, J.~Men\'endez, and F.~Vissani, ``{Toward the discovery of matter creation with neutrinoless \ensuremath{\beta}\ensuremath{\beta} decay},'' \href{http://dx.doi.org/10.1103/RevModPhys.95.025002}{{\em Rev. Mod. Phys.} {\bfseries 95} no.~2, (2023) 025002}, \href{http://arxiv.org/abs/2202.01787}{{\ttfamily arXiv:2202.01787 [hep-ex]}}.

\bibitem{GERDA:2020xhi}
{\bfseries GERDA} Collaboration, M.~Agostini {\em et~al.}, ``{Final Results of GERDA on the Search for Neutrinoless Double-$\beta$ Decay},'' \href{http://dx.doi.org/10.1103/PhysRevLett.125.252502}{{\em Phys. Rev. Lett.} {\bfseries 125} no.~25, (2020) 252502}, \href{http://arxiv.org/abs/2009.06079}{{\ttfamily arXiv:2009.06079 [nucl-ex]}}.

\bibitem{KamLAND-Zen:2022tow}
{\bfseries KamLAND-Zen} Collaboration, S.~Abe {\em et~al.}, ``{Search for the Majorana Nature of Neutrinos in the Inverted Mass Ordering Region with KamLAND-Zen},'' \href{http://dx.doi.org/10.1103/PhysRevLett.130.051801}{{\em Phys. Rev. Lett.} {\bfseries 130} no.~5, (2023) 051801}, \href{http://arxiv.org/abs/2203.02139}{{\ttfamily arXiv:2203.02139 [hep-ex]}}.

\bibitem{LEGEND:2021bnm}
{\bfseries LEGEND} Collaboration, N.~Abgrall {\em et~al.}, ``{The Large Enriched Germanium Experiment for Neutrinoless $\beta\beta$ Decay}: {LEGEND-1000 Preconceptual Design Report},'' \href{http://arxiv.org/abs/2107.11462}{{\ttfamily arXiv:2107.11462 [physics.ins-det]}}.

\bibitem{nEXO:2021ujk}
{\bfseries nEXO} Collaboration, G.~Adhikari {\em et~al.}, ``{nEXO: neutrinoless double beta decay search beyond 10$^{28}$ year half-life sensitivity},'' \href{http://dx.doi.org/10.1088/1361-6471/ac3631}{{\em J. Phys. G} {\bfseries 49} no.~1, (2022) 015104}, \href{http://arxiv.org/abs/2106.16243}{{\ttfamily arXiv:2106.16243 [nucl-ex]}}.

\bibitem{Esteban:2024eli}
I.~Esteban, M.~C. Gonzalez-Garcia, M.~Maltoni, I.~Martinez-Soler, J.~a.~P. Pinheiro, and T.~Schwetz, ``{NuFit-6.0: updated global analysis of three-flavor neutrino oscillations},'' \href{http://dx.doi.org/10.1007/JHEP12(2024)216}{{\em JHEP} {\bfseries 12} (2024) 216}, \href{http://arxiv.org/abs/2410.05380}{{\ttfamily arXiv:2410.05380 [hep-ph]}}.

\bibitem{DESI:2024mwx}
{\bfseries DESI} Collaboration, A.~G. Adame {\em et~al.}, ``{DESI 2024 VI: cosmological constraints from the measurements of baryon acoustic oscillations},'' \href{http://dx.doi.org/10.1088/1475-7516/2025/02/021}{{\em JCAP} {\bfseries 02} (2025) 021}, \href{http://arxiv.org/abs/2404.03002}{{\ttfamily arXiv:2404.03002 [astro-ph.CO]}}.

\bibitem{Deppisch:2015qwa}
F.~F. Deppisch, P.~S.~B. Dev, and A.~Pilaftsis, ``{Neutrinos and Collider Physics},'' \href{http://dx.doi.org/10.1088/1367-2630/17/7/075019}{{\em New J. Phys.} {\bfseries 17} no.~7, (2015) 075019}, \href{http://arxiv.org/abs/1502.06541}{{\ttfamily arXiv:1502.06541 [hep-ph]}}.

\bibitem{Cai:2017mow}
Y.~Cai, T.~Han, T.~Li, and R.~Ruiz, ``{Lepton Number Violation: Seesaw Models and Their Collider Tests},'' \href{http://dx.doi.org/10.3389/fphy.2018.00040}{{\em Front. in Phys.} {\bfseries 6} (2018) 40}, \href{http://arxiv.org/abs/1711.02180}{{\ttfamily arXiv:1711.02180 [hep-ph]}}.

\bibitem{Weinberg:1979sa}
S.~Weinberg, ``{Baryon and Lepton Nonconserving Processes},'' \href{http://dx.doi.org/10.1103/PhysRevLett.43.1566}{{\em Phys. Rev. Lett.} {\bfseries 43} (1979) 1566--1570}.

\bibitem{Minkowski:1977sc}
P.~Minkowski, ``{$\mu \to e\gamma$ at a Rate of One Out of $10^{9}$ Muon Decays?},'' \href{http://dx.doi.org/10.1016/0370-2693(77)90435-X}{{\em Phys. Lett. B} {\bfseries 67} (1977) 421--428}.

\bibitem{Mohapatra:1979ia}
R.~N. Mohapatra and G.~Senjanovic, ``{Neutrino Mass and Spontaneous Parity Nonconservation},'' \href{http://dx.doi.org/10.1103/PhysRevLett.44.912}{{\em Phys. Rev. Lett.} {\bfseries 44} (1980) 912}.

\bibitem{Yanagida:1979as}
T.~Yanagida, ``{Horizontal gauge symmetry and masses of neutrinos},'' {\em Conf. Proc. C} {\bfseries 7902131} (1979) 95--99.

\bibitem{Gell-Mann:1979vob}
M.~Gell-Mann, P.~Ramond, and R.~Slansky, ``{Complex Spinors and Unified Theories},'' {\em Conf. Proc. C} {\bfseries 790927} (1979) 315--321, \href{http://arxiv.org/abs/1306.4669}{{\ttfamily arXiv:1306.4669 [hep-th]}}.

\bibitem{Schechter:1980gr}
J.~Schechter and J.~W.~F. Valle, ``{Neutrino Masses in SU(2) x U(1) Theories},'' \href{http://dx.doi.org/10.1103/PhysRevD.22.2227}{{\em Phys. Rev. D} {\bfseries 22} (1980) 2227}.

\bibitem{Magg:1980ut}
M.~Magg and C.~Wetterich, ``{Neutrino Mass Problem and Gauge Hierarchy},'' \href{http://dx.doi.org/10.1016/0370-2693(80)90825-4}{{\em Phys. Lett. B} {\bfseries 94} (1980) 61--64}.

\bibitem{Cheng:1980qt}
T.~P. Cheng and L.-F. Li, ``{Neutrino Masses, Mixings and Oscillations in SU(2) x U(1) Models of Electroweak Interactions},'' \href{http://dx.doi.org/10.1103/PhysRevD.22.2860}{{\em Phys. Rev. D} {\bfseries 22} (1980) 2860}.

\bibitem{Lazarides:1980nt}
G.~Lazarides, Q.~Shafi, and C.~Wetterich, ``{Proton Lifetime and Fermion Masses in an SO(10) Model},'' \href{http://dx.doi.org/10.1016/0550-3213(81)90354-0}{{\em Nucl. Phys. B} {\bfseries 181} (1981) 287--300}.

\bibitem{Mohapatra:1980yp}
R.~N. Mohapatra and G.~Senjanovic, ``{Neutrino Masses and Mixings in Gauge Models with Spontaneous Parity Violation},'' \href{http://dx.doi.org/10.1103/PhysRevD.23.165}{{\em Phys. Rev. D} {\bfseries 23} (1981) 165}.

\bibitem{Foot:1988aq}
R.~Foot, H.~Lew, X.~G. He, and G.~C. Joshi, ``{Seesaw Neutrino Masses Induced by a Triplet of Leptons},'' \href{http://dx.doi.org/10.1007/BF01415558}{{\em Z. Phys. C} {\bfseries 44} (1989) 441}.

\bibitem{Keung:1983uu}
W.-Y. Keung and G.~Senjanovic, ``{Majorana Neutrinos and the Production of the Right-handed Charged Gauge Boson},'' \href{http://dx.doi.org/10.1103/PhysRevLett.50.1427}{{\em Phys. Rev. Lett.} {\bfseries 50} (1983) 1427}.

\bibitem{Kersten:2007vk}
J.~Kersten and A.~Y. Smirnov, ``{Right-Handed Neutrinos at CERN LHC and the Mechanism of Neutrino Mass Generation},'' \href{http://dx.doi.org/10.1103/PhysRevD.76.073005}{{\em Phys. Rev. D} {\bfseries 76} (2007) 073005}, \href{http://arxiv.org/abs/0705.3221}{{\ttfamily arXiv:0705.3221 [hep-ph]}}.

\bibitem{Mohapatra:1986bd}
R.~N. Mohapatra and J.~W.~F. Valle, ``{Neutrino Mass and Baryon Number Nonconservation in Superstring Models},'' \href{http://dx.doi.org/10.1103/PhysRevD.34.1642}{{\em Phys. Rev. D} {\bfseries 34} (1986) 1642}.

\bibitem{Malinsky:2005bi}
M.~Malinsky, J.~C. Romao, and J.~W.~F. Valle, ``{Novel supersymmetric SO(10) seesaw mechanism},'' \href{http://dx.doi.org/10.1103/PhysRevLett.95.161801}{{\em Phys. Rev. Lett.} {\bfseries 95} (2005) 161801}, \href{http://arxiv.org/abs/hep-ph/0506296}{{\ttfamily arXiv:hep-ph/0506296}}.

\bibitem{Dev:2012sg}
P.~S.~B. Dev and A.~Pilaftsis, ``{Minimal Radiative Neutrino Mass Mechanism for Inverse Seesaw Models},'' \href{http://dx.doi.org/10.1103/PhysRevD.86.113001}{{\em Phys. Rev. D} {\bfseries 86} (2012) 113001}, \href{http://arxiv.org/abs/1209.4051}{{\ttfamily arXiv:1209.4051 [hep-ph]}}.

\bibitem{Bolton:2019pcu}
P.~D. Bolton, F.~F. Deppisch, and P.~S.~B. Dev, ``{Neutrinoless double beta decay versus other probes of heavy sterile neutrinos},'' \href{http://dx.doi.org/10.1007/JHEP03(2020)170}{{\em JHEP} {\bfseries 03} (2020) 170}, \href{http://arxiv.org/abs/1912.03058}{{\ttfamily arXiv:1912.03058 [hep-ph]}}. For updates, see \url{www.sterile-neutrino.org}.

\bibitem{Bolton:2022pyf}
P.~D. Bolton, F.~F. Deppisch, and P.~S.~B. Dev, ``{Probes of Heavy Sterile Neutrinos},'' in {\em {56th Rencontres de Moriond on Electroweak Interactions and Unified Theories}}.
\newblock 6, 2022.
\newblock \href{http://arxiv.org/abs/2206.01140}{{\ttfamily arXiv:2206.01140 [hep-ph]}}.

\bibitem{Tastet:2021vwp}
J.-L. Tastet, O.~Ruchayskiy, and I.~Timiryasov, ``{Reinterpreting the ATLAS bounds on heavy neutral leptons in a realistic neutrino oscillation model},'' \href{http://dx.doi.org/10.1007/JHEP12(2021)182}{{\em JHEP} {\bfseries 12} (2021) 182}, \href{http://arxiv.org/abs/2107.12980}{{\ttfamily arXiv:2107.12980 [hep-ph]}}.

\bibitem{Antusch:2016ejd}
S.~Antusch, E.~Cazzato, and O.~Fischer, ``{Sterile neutrino searches at future $e^-e^+$, $pp$, and $e^-p$ colliders},'' \href{http://dx.doi.org/10.1142/S0217751X17500786}{{\em Int. J. Mod. Phys. A} {\bfseries 32} no.~14, (2017) 1750078}, \href{http://arxiv.org/abs/1612.02728}{{\ttfamily arXiv:1612.02728 [hep-ph]}}.

\bibitem{Abdullahi:2022jlv}
A.~M. Abdullahi {\em et~al.}, ``{The present and future status of heavy neutral leptons},'' \href{http://dx.doi.org/10.1088/1361-6471/ac98f9}{{\em J. Phys. G} {\bfseries 50} no.~2, (2023) 020501}, \href{http://arxiv.org/abs/2203.08039}{{\ttfamily arXiv:2203.08039 [hep-ph]}}.

\bibitem{Dasgupta:2021ies}
B.~Dasgupta and J.~Kopp, ``{Sterile Neutrinos},'' \href{http://dx.doi.org/10.1016/j.physrep.2021.06.002}{{\em Phys. Rept.} {\bfseries 928} (2021) 1--63}, \href{http://arxiv.org/abs/2106.05913}{{\ttfamily arXiv:2106.05913 [hep-ph]}}.

\bibitem{Dev:2012zg}
P.~S.~B. Dev, R.~Franceschini, and R.~N. Mohapatra, ``{Bounds on TeV Seesaw Models from LHC Higgs Data},'' \href{http://dx.doi.org/10.1103/PhysRevD.86.093010}{{\em Phys. Rev. D} {\bfseries 86} (2012) 093010}, \href{http://arxiv.org/abs/1207.2756}{{\ttfamily arXiv:1207.2756 [hep-ph]}}.

\bibitem{Das:2017zjc}
A.~Das, P.~S.~B. Dev, and C.~S. Kim, ``{Constraining Sterile Neutrinos from Precision Higgs Data},'' \href{http://dx.doi.org/10.1103/PhysRevD.95.115013}{{\em Phys. Rev. D} {\bfseries 95} no.~11, (2017) 115013}, \href{http://arxiv.org/abs/1704.00880}{{\ttfamily arXiv:1704.00880 [hep-ph]}}.

\bibitem{Gao:2021one}
Y.~Gao and K.~Wang, ``{Heavy neutrino searches via same-sign lepton pairs at a Higgs boson factory},'' \href{http://dx.doi.org/10.1103/PhysRevD.105.076005}{{\em Phys. Rev. D} {\bfseries 105} no.~7, (2022) 076005}, \href{http://arxiv.org/abs/2102.12826}{{\ttfamily arXiv:2102.12826 [hep-ph]}}.

\bibitem{ATLAS:2019kpx}
{\bfseries ATLAS} Collaboration, G.~Aad {\em et~al.}, ``{Search for heavy neutral leptons in decays of $W$ bosons produced in 13 TeV $pp$ collisions using prompt and displaced signatures with the ATLAS detector},'' \href{http://dx.doi.org/10.1007/JHEP10(2019)265}{{\em JHEP} {\bfseries 10} (2019) 265}, \href{http://arxiv.org/abs/1905.09787}{{\ttfamily arXiv:1905.09787 [hep-ex]}}.

\bibitem{CMS:2024xdq}
{\bfseries CMS} Collaboration, A.~Hayrapetyan {\em et~al.}, ``{Search for heavy neutral leptons in final states with electrons, muons, and hadronically decaying tau leptons in proton-proton collisions at $ \sqrt{s} $ = 13 TeV},'' \href{http://dx.doi.org/10.1007/JHEP06(2024)123}{{\em JHEP} {\bfseries 06} (2024) 123}, \href{http://arxiv.org/abs/2403.00100}{{\ttfamily arXiv:2403.00100 [hep-ex]}}.

\bibitem{CMS:2024hik}
{\bfseries CMS} Collaboration, A.~Hayrapetyan {\em et~al.}, ``{Search for long-lived heavy neutral leptons in proton-proton collision events with a lepton-jet pair associated with a secondary vertex at $ \sqrt{s} $ = 13 TeV},'' \href{http://dx.doi.org/10.1007/JHEP02(2025)036}{{\em JHEP} {\bfseries 02} (2025) 036}, \href{http://arxiv.org/abs/2407.10717}{{\ttfamily arXiv:2407.10717 [hep-ex]}}.

\bibitem{ATLAS:2025uah}
{\bfseries ATLAS} Collaboration, G.~Aad {\em et~al.}, ``{Search for heavy neutral leptons in decays of W bosons using leptonic and semi-leptonic displaced vertices in $\sqrt{s} = 13$ TeV $pp$ collisions with the ATLAS detector},'' \href{http://arxiv.org/abs/2503.16213}{{\ttfamily arXiv:2503.16213 [hep-ex]}}.

\bibitem{Blondel:2014bra}
{\bfseries FCC-ee study Team} Collaboration, A.~Blondel, E.~Graverini, N.~Serra, and M.~Shaposhnikov, ``{Search for Heavy Right Handed Neutrinos at the FCC-ee},'' \href{http://dx.doi.org/10.1016/j.nuclphysbps.2015.09.304}{{\em Nucl. Part. Phys. Proc.} {\bfseries 273-275} (2016) 1883--1890}, \href{http://arxiv.org/abs/1411.5230}{{\ttfamily arXiv:1411.5230 [hep-ex]}}.

\bibitem{Li:2023tbx}
P.~Li, Z.~Liu, and K.-F. Lyu, ``{Heavy neutral leptons at muon colliders},'' \href{http://dx.doi.org/10.1007/JHEP03(2023)231}{{\em JHEP} {\bfseries 03} (2023) 231}, \href{http://arxiv.org/abs/2301.07117}{{\ttfamily arXiv:2301.07117 [hep-ph]}}.

\bibitem{Bellagamba:2025xpd}
L.~Bellagamba, G.~Polesello, and N.~Valle, ``{Searches for Heavy Neutral Leptons at FCC-ee in final states including a muon},'' \href{http://arxiv.org/abs/2503.19464}{{\ttfamily arXiv:2503.19464 [hep-ex]}}.

\bibitem{Fernandez-Martinez:2015hxa}
E.~Fernandez-Martinez, J.~Hernandez-Garcia, J.~Lopez-Pavon, and M.~Lucente, ``{Loop level constraints on Seesaw neutrino mixing},'' \href{http://dx.doi.org/10.1007/JHEP10(2015)130}{{\em JHEP} {\bfseries 10} (2015) 130}, \href{http://arxiv.org/abs/1508.03051}{{\ttfamily arXiv:1508.03051 [hep-ph]}}.

\bibitem{Bray:2007ru}
S.~Bray, J.~S. Lee, and A.~Pilaftsis, ``{Resonant CP violation due to heavy neutrinos at the LHC},'' \href{http://dx.doi.org/10.1016/j.nuclphysb.2007.07.002}{{\em Nucl. Phys. B} {\bfseries 786} (2007) 95--118}, \href{http://arxiv.org/abs/hep-ph/0702294}{{\ttfamily arXiv:hep-ph/0702294}}.

\bibitem{Anamiati:2016uxp}
G.~Anamiati, M.~Hirsch, and E.~Nardi, ``{Quasi-Dirac neutrinos at the LHC},'' \href{http://dx.doi.org/10.1007/JHEP10(2016)010}{{\em JHEP} {\bfseries 10} (2016) 010}, \href{http://arxiv.org/abs/1607.05641}{{\ttfamily arXiv:1607.05641 [hep-ph]}}.

\bibitem{Antusch:2017ebe}
S.~Antusch, E.~Cazzato, and O.~Fischer, ``{Resolvable heavy neutrino\textendash{}antineutrino oscillations at colliders},'' \href{http://dx.doi.org/10.1142/S0217732319500615}{{\em Mod. Phys. Lett. A} {\bfseries 34} no.~07n08, (2019) 1950061}, \href{http://arxiv.org/abs/1709.03797}{{\ttfamily arXiv:1709.03797 [hep-ph]}}.

\bibitem{Antusch:2024otj}
S.~Antusch, J.~Hajer, and B.~M.~S. Oliveira, ``{Discovering heavy neutrino-antineutrino oscillations at the Z-pole},'' \href{http://dx.doi.org/10.1007/JHEP11(2024)102}{{\em JHEP} {\bfseries 11} (2024) 102}, \href{http://arxiv.org/abs/2408.01389}{{\ttfamily arXiv:2408.01389 [hep-ph]}}.

\bibitem{Deppisch:2013cya}
F.~F. Deppisch, N.~Desai, and J.~W.~F. Valle, ``{Is charged lepton flavor violation a high energy phenomenon?},'' \href{http://dx.doi.org/10.1103/PhysRevD.89.051302}{{\em Phys. Rev. D} {\bfseries 89} no.~5, (2014) 051302}, \href{http://arxiv.org/abs/1308.6789}{{\ttfamily arXiv:1308.6789 [hep-ph]}}.

\bibitem{Das:2019fee}
A.~Das, P.~S.~B. Dev, and N.~Okada, ``{Long-lived TeV-scale right-handed neutrino production at the LHC in gauged $U(1)_X$ model},'' \href{http://dx.doi.org/10.1016/j.physletb.2019.135052}{{\em Phys. Lett. B} {\bfseries 799} (2019) 135052}, \href{http://arxiv.org/abs/1906.04132}{{\ttfamily arXiv:1906.04132 [hep-ph]}}.

\bibitem{Das:2021esm}
A.~Das, P.~S.~B. Dev, Y.~Hosotani, and S.~Mandal, ``{Probing the minimal U(1)X model at future electron-positron colliders via fermion pair-production channels},'' \href{http://dx.doi.org/10.1103/PhysRevD.105.115030}{{\em Phys. Rev. D} {\bfseries 105} no.~11, (2022) 115030}, \href{http://arxiv.org/abs/2104.10902}{{\ttfamily arXiv:2104.10902 [hep-ph]}}.

\bibitem{Das:2017flq}
A.~Das, N.~Okada, and D.~Raut, ``{Enhanced pair production of heavy Majorana neutrinos at the LHC},'' \href{http://dx.doi.org/10.1103/PhysRevD.97.115023}{{\em Phys. Rev. D} {\bfseries 97} no.~11, (2018) 115023}, \href{http://arxiv.org/abs/1710.03377}{{\ttfamily arXiv:1710.03377 [hep-ph]}}.

\bibitem{Dasgupta:2023zrh}
A.~Dasgupta, P.~S.~B. Dev, T.~Han, R.~Padhan, S.~Wang, and K.~Xie, ``{Searching for heavy leptophilic Z': from lepton colliders to gravitational waves},'' \href{http://dx.doi.org/10.1007/JHEP12(2023)011}{{\em JHEP} {\bfseries 12} (2023) 011}, \href{http://arxiv.org/abs/2308.12804}{{\ttfamily arXiv:2308.12804 [hep-ph]}}.

\bibitem{Chen:2013foz}
C.-Y. Chen, P.~S.~B. Dev, and R.~N. Mohapatra, ``{Probing Heavy-Light Neutrino Mixing in Left-Right Seesaw Models at the LHC},'' \href{http://dx.doi.org/10.1103/PhysRevD.88.033014}{{\em Phys. Rev. D} {\bfseries 88} (2013) 033014}, \href{http://arxiv.org/abs/1306.2342}{{\ttfamily arXiv:1306.2342 [hep-ph]}}.

\bibitem{Nemevsek:2018bbt}
M.~Nemev\v{s}ek, F.~Nesti, and G.~Popara, ``{Keung-Senjanovi\'c process at the LHC: From lepton number violation to displaced vertices to invisible decays},'' \href{http://dx.doi.org/10.1103/PhysRevD.97.115018}{{\em Phys. Rev. D} {\bfseries 97} no.~11, (2018) 115018}, \href{http://arxiv.org/abs/1801.05813}{{\ttfamily arXiv:1801.05813 [hep-ph]}}.

\bibitem{Biswal:2017nfl}
S.~S. Biswal and P.~S.~B. Dev, ``{Probing left-right seesaw models using beam polarization at an $e^+e^-$ collider},'' \href{http://dx.doi.org/10.1103/PhysRevD.95.115031}{{\em Phys. Rev. D} {\bfseries 95} no.~11, (2017) 115031}, \href{http://arxiv.org/abs/1701.08751}{{\ttfamily arXiv:1701.08751 [hep-ph]}}.

\bibitem{FileviezPerez:2008jbu}
P.~Fileviez~Perez, T.~Han, G.-y. Huang, T.~Li, and K.~Wang, ``{Neutrino Masses and the CERN LHC: Testing Type II Seesaw},'' \href{http://dx.doi.org/10.1103/PhysRevD.78.015018}{{\em Phys. Rev. D} {\bfseries 78} (2008) 015018}, \href{http://arxiv.org/abs/0805.3536}{{\ttfamily arXiv:0805.3536 [hep-ph]}}.

\bibitem{CMS:2017pet}
{\bfseries CMS} Collaboration, ``{A search for doubly-charged Higgs boson production in three and four lepton final states at $\sqrt{s}=13~\mathrm{TeV}$},''.

\bibitem{ATLAS:2021jol}
{\bfseries ATLAS} Collaboration, G.~Aad {\em et~al.}, ``{Search for doubly and singly charged Higgs bosons decaying into vector bosons in multi-lepton final states with the ATLAS detector using proton-proton collisions at $ \sqrt{\mathrm{s}} $ = 13 TeV},'' \href{http://dx.doi.org/10.1007/JHEP06(2021)146}{{\em JHEP} {\bfseries 06} (2021) 146}, \href{http://arxiv.org/abs/2101.11961}{{\ttfamily arXiv:2101.11961 [hep-ex]}}.

\bibitem{ATLAS:2022pbd}
{\bfseries ATLAS} Collaboration, G.~Aad {\em et~al.}, ``{Search for doubly charged Higgs boson production in multi-lepton final states using 139~fb$^{-1}$ of proton\textendash{}proton collisions at $\sqrt{s}$ = 13~TeV with the ATLAS detector},'' \href{http://dx.doi.org/10.1140/epjc/s10052-023-11578-9}{{\em Eur. Phys. J. C} {\bfseries 83} no.~7, (2023) 605}, \href{http://arxiv.org/abs/2211.07505}{{\ttfamily arXiv:2211.07505 [hep-ex]}}.

\bibitem{Gunion:1989in}
J.~F. Gunion, J.~Grifols, A.~Mendez, B.~Kayser, and F.~I. Olness, ``{Higgs Bosons in Left-Right Symmetric Models},'' \href{http://dx.doi.org/10.1103/PhysRevD.40.1546}{{\em Phys. Rev. D} {\bfseries 40} (1989) 1546}.

\bibitem{Dev:2016dja}
P.~S.~B. Dev, R.~N. Mohapatra, and Y.~Zhang, ``{Probing the Higgs Sector of the Minimal Left-Right Symmetric Model at Future Hadron Colliders},'' \href{http://dx.doi.org/10.1007/JHEP05(2016)174}{{\em JHEP} {\bfseries 05} (2016) 174}, \href{http://arxiv.org/abs/1602.05947}{{\ttfamily arXiv:1602.05947 [hep-ph]}}.

\bibitem{Dev:2016nfr}
P.~S.~B. Dev, R.~N. Mohapatra, and Y.~Zhang, ``{Displaced photon signal from a possible light scalar in minimal left-right seesaw model},'' \href{http://dx.doi.org/10.1103/PhysRevD.95.115001}{{\em Phys. Rev. D} {\bfseries 95} no.~11, (2017) 115001}, \href{http://arxiv.org/abs/1612.09587}{{\ttfamily arXiv:1612.09587 [hep-ph]}}.

\bibitem{Dev:2017dui}
P.~S.~B. Dev, R.~N. Mohapatra, and Y.~Zhang, ``{Long Lived Light Scalars as Probe of Low Scale Seesaw Models},'' \href{http://dx.doi.org/10.1016/j.nuclphysb.2017.07.021}{{\em Nucl. Phys. B} {\bfseries 923} (2017) 179--221}, \href{http://arxiv.org/abs/1703.02471}{{\ttfamily arXiv:1703.02471 [hep-ph]}}.

\bibitem{Maiezza:2015lza}
A.~Maiezza, M.~Nemev\v{s}ek, and F.~Nesti, ``{Lepton Number Violation in Higgs Decay at LHC},'' \href{http://dx.doi.org/10.1103/PhysRevLett.115.081802}{{\em Phys. Rev. Lett.} {\bfseries 115} (2015) 081802}, \href{http://arxiv.org/abs/1503.06834}{{\ttfamily arXiv:1503.06834 [hep-ph]}}.

\bibitem{Nemevsek:2016enw}
M.~Nemev\v{s}ek, F.~Nesti, and J.~C. Vasquez, ``{Majorana Higgses at colliders},'' \href{http://dx.doi.org/10.1007/JHEP04(2017)114}{{\em JHEP} {\bfseries 04} (2017) 114}, \href{http://arxiv.org/abs/1612.06840}{{\ttfamily arXiv:1612.06840 [hep-ph]}}.

\bibitem{Cai:2017jrq}
Y.~Cai, J.~Herrero-Garc\'\i{}a, M.~A. Schmidt, A.~Vicente, and R.~R. Volkas, ``{From the trees to the forest: a review of radiative neutrino mass models},'' \href{http://dx.doi.org/10.3389/fphy.2017.00063}{{\em Front. in Phys.} {\bfseries 5} (2017) 63}, \href{http://arxiv.org/abs/1706.08524}{{\ttfamily arXiv:1706.08524 [hep-ph]}}.

\bibitem{Zee:1980ai}
A.~Zee, ``{A Theory of Lepton Number Violation, Neutrino Majorana Mass, and Oscillation},'' \href{http://dx.doi.org/10.1016/0370-2693(80)90349-4}{{\em Phys. Lett. B} {\bfseries 93} (1980) 389}. [Erratum: Phys.Lett.B 95, 461 (1980)].

\bibitem{Zee:1985id}
A.~Zee, ``{Quantum Numbers of Majorana Neutrino Masses},'' \href{http://dx.doi.org/10.1016/0550-3213(86)90475-X}{{\em Nucl. Phys. B} {\bfseries 264} (1986) 99}.

\bibitem{Babu:1988ki}
K.~S. Babu, ``{Model of 'Calculable' Majorana Neutrino Masses},'' \href{http://dx.doi.org/10.1016/0370-2693(88)91584-5}{{\em Phys. Lett. B} {\bfseries 203} (1988) 132}.

\bibitem{Krauss:2002px}
L.~M. Krauss, S.~Nasri, and M.~Trodden, ``{A Model for neutrino masses and dark matter},'' \href{http://dx.doi.org/10.1103/PhysRevD.67.085002}{{\em Phys. Rev. D} {\bfseries 67} (2003) 085002}, \href{http://arxiv.org/abs/hep-ph/0210389}{{\ttfamily arXiv:hep-ph/0210389}}.

\bibitem{ALEPH:2013htx}
{\bfseries ALEPH, DELPHI, L3, OPAL, LEP} Collaboration, G.~Abbiendi {\em et~al.}, ``{Search for Charged Higgs bosons: Combined Results Using LEP Data},'' \href{http://dx.doi.org/10.1140/epjc/s10052-013-2463-1}{{\em Eur. Phys. J. C} {\bfseries 73} (2013) 2463}, \href{http://arxiv.org/abs/1301.6065}{{\ttfamily arXiv:1301.6065 [hep-ex]}}.

\bibitem{Babu:2019mfe}
K.~S. Babu, P.~S.~B. Dev, S.~Jana, and A.~Thapa, ``{Non-Standard Interactions in Radiative Neutrino Mass Models},'' \href{http://dx.doi.org/10.1007/JHEP03(2020)006}{{\em JHEP} {\bfseries 03} (2020) 006}, \href{http://arxiv.org/abs/1907.09498}{{\ttfamily arXiv:1907.09498 [hep-ph]}}.

\bibitem{Dev:2023nha}
P.~S.~B. Dev, J.~Heeck, and A.~Thapa, ``{Neutrino mass models at $\mu $TRISTAN},'' \href{http://dx.doi.org/10.1140/epjc/s10052-024-12496-0}{{\em Eur. Phys. J. C} {\bfseries 84} no.~2, (2024) 148}, \href{http://arxiv.org/abs/2309.06463}{{\ttfamily arXiv:2309.06463 [hep-ph]}}.

\bibitem{Buonocore:2020erb}
L.~Buonocore, U.~Haisch, P.~Nason, F.~Tramontano, and G.~Zanderighi, ``{Lepton-Quark Collisions at the Large Hadron Collider},'' \href{http://dx.doi.org/10.1103/PhysRevLett.125.231804}{{\em Phys. Rev. Lett.} {\bfseries 125} no.~23, (2020) 231804}, \href{http://arxiv.org/abs/2005.06475}{{\ttfamily arXiv:2005.06475 [hep-ph]}}.

\bibitem{Afik:2023vyl}
Y.~Afik, P.~S. Bhupal~Dev, and A.~Thapa, ``{Hints of a new leptophilic Higgs sector?},'' \href{http://dx.doi.org/10.1103/PhysRevD.109.015003}{{\em Phys. Rev. D} {\bfseries 109} no.~1, (2024) 015003}, \href{http://arxiv.org/abs/2305.19314}{{\ttfamily arXiv:2305.19314 [hep-ph]}}.

\bibitem{Fukugita:1986hr}
M.~Fukugita and T.~Yanagida, ``{Baryogenesis Without Grand Unification},'' \href{http://dx.doi.org/10.1016/0370-2693(86)91126-3}{{\em Phys. Lett. B} {\bfseries 174} (1986) 45--47}.

\bibitem{Bodeker:2020ghk}
D.~Bodeker and W.~Buchmuller, ``{Baryogenesis from the weak scale to the grand unification scale},'' \href{http://dx.doi.org/10.1103/RevModPhys.93.035004}{{\em Rev. Mod. Phys.} {\bfseries 93} no.~3, (2021) 035004}, \href{http://arxiv.org/abs/2009.07294}{{\ttfamily arXiv:2009.07294 [hep-ph]}}.

\bibitem{Pilaftsis:2003gt}
A.~Pilaftsis and T.~E.~J. Underwood, ``{Resonant leptogenesis},'' \href{http://dx.doi.org/10.1016/j.nuclphysb.2004.05.029}{{\em Nucl. Phys. B} {\bfseries 692} (2004) 303--345}, \href{http://arxiv.org/abs/hep-ph/0309342}{{\ttfamily arXiv:hep-ph/0309342}}.

\bibitem{Akhmedov:1998qx}
E.~K. Akhmedov, V.~A. Rubakov, and A.~Y. Smirnov, ``{Baryogenesis via neutrino oscillations},'' \href{http://dx.doi.org/10.1103/PhysRevLett.81.1359}{{\em Phys. Rev. Lett.} {\bfseries 81} (1998) 1359--1362}, \href{http://arxiv.org/abs/hep-ph/9803255}{{\ttfamily arXiv:hep-ph/9803255}}.

\bibitem{Klaric:2020phc}
J.~Klari\'c, M.~Shaposhnikov, and I.~Timiryasov, ``{Uniting Low-Scale Leptogenesis Mechanisms},'' \href{http://dx.doi.org/10.1103/PhysRevLett.127.111802}{{\em Phys. Rev. Lett.} {\bfseries 127} no.~11, (2021) 111802}, \href{http://arxiv.org/abs/2008.13771}{{\ttfamily arXiv:2008.13771 [hep-ph]}}.

\bibitem{Drewes:2021nqr}
M.~Drewes, Y.~Georis, and J.~Klari\'c, ``{Mapping the Viable Parameter Space for Testable Leptogenesis},'' \href{http://dx.doi.org/10.1103/PhysRevLett.128.051801}{{\em Phys. Rev. Lett.} {\bfseries 128} no.~5, (2022) 051801}, \href{http://arxiv.org/abs/2106.16226}{{\ttfamily arXiv:2106.16226 [hep-ph]}}.

\bibitem{Drewes:2016upu}
M.~Drewes {\em et~al.}, ``{A White Paper on keV Sterile Neutrino Dark Matter},'' \href{http://dx.doi.org/10.1088/1475-7516/2017/01/025}{{\em JCAP} {\bfseries 01} (2017) 025}, \href{http://arxiv.org/abs/1602.04816}{{\ttfamily arXiv:1602.04816 [hep-ph]}}.

\bibitem{Dodelson:1993je}
S.~Dodelson and L.~M. Widrow, ``{Sterile-neutrinos as dark matter},'' \href{http://dx.doi.org/10.1103/PhysRevLett.72.17}{{\em Phys. Rev. Lett.} {\bfseries 72} (1994) 17--20}, \href{http://arxiv.org/abs/hep-ph/9303287}{{\ttfamily arXiv:hep-ph/9303287}}.

\bibitem{DeGouvea:2019wpf}
A.~De~Gouv\^ea, M.~Sen, W.~Tangarife, and Y.~Zhang, ``{Dodelson-Widrow Mechanism in the Presence of Self-Interacting Neutrinos},'' \href{http://dx.doi.org/10.1103/PhysRevLett.124.081802}{{\em Phys. Rev. Lett.} {\bfseries 124} no.~8, (2020) 081802}, \href{http://arxiv.org/abs/1910.04901}{{\ttfamily arXiv:1910.04901 [hep-ph]}}.

\bibitem{Astros:2023xhe}
M.~D. Astros and S.~Vogl, ``{Boosting the production of sterile neutrino dark matter with self-interactions},'' \href{http://dx.doi.org/10.1007/JHEP03(2024)032}{{\em JHEP} {\bfseries 03} (2024) 032}, \href{http://arxiv.org/abs/2307.15565}{{\ttfamily arXiv:2307.15565 [hep-ph]}}.

\bibitem{Nemevsek:2012cd}
M.~Nemevsek, G.~Senjanovic, and Y.~Zhang, ``{Warm Dark Matter in Low Scale Left-Right Theory},'' \href{http://dx.doi.org/10.1088/1475-7516/2012/07/006}{{\em JCAP} {\bfseries 07} (2012) 006}, \href{http://arxiv.org/abs/1205.0844}{{\ttfamily arXiv:1205.0844 [hep-ph]}}.

\bibitem{Dev:2025fcv}
P.~S.~B. Dev, J.~Heeck, and A.~Thapa, ``{Decaying scalar dark matter in the minimal left-right symmetric model},'' \href{http://arxiv.org/abs/2501.14669}{{\ttfamily arXiv:2501.14669 [hep-ph]}}.

\bibitem{Abazajian:2017tcc}
K.~N. Abazajian, ``{Sterile neutrinos in cosmology},'' \href{http://dx.doi.org/10.1016/j.physrep.2017.10.003}{{\em Phys. Rept.} {\bfseries 711-712} (2017) 1--28}, \href{http://arxiv.org/abs/1705.01837}{{\ttfamily arXiv:1705.01837 [hep-ph]}}.

\end{thebibliography}\endgroup
\end{document}